\documentclass[a4paper, twocolumn, 11pt,accepted=2025-01-28]{quantumarticle}
\pdfoutput=1
\usepackage[utf8]{inputenc}
\usepackage[english]{babel}
\usepackage[T1]{fontenc}
\usepackage[numbers,sort&compress]{natbib}
\usepackage{amsmath, amssymb}
\usepackage{enumitem}
\usepackage{xcolor}
\usepackage{graphicx}
\usepackage[none]{hyphenat}
\usepackage[export]{adjustbox}
\usepackage[colorlinks]{hyperref}
\begin{document}
\title{Spectral chaos bounds from scaling theory of maximally efficient quantum-dynamical scrambling}
\author{Tara Kalsi}
\email{t.kalsi@lancaster.ac.uk}
\orcid{0000-0003-1513-5795}
\author{Alessandro Romito}
\orcid{0000-0003-3082-6279}
\author{Henning Schomerus}
\orcid{0000-0002-7959-0992}
\affiliation{Department of Physics, Lancaster University, Lancaster LA1 4YB, United Kingdom}
\maketitle
\begin{abstract}
A key conjecture about the evolution of complex quantum systems towards an ergodic steady state, known as scrambling, is that this process acquires universal features when it is most efficient. We develop a single-parameter scaling theory for the spectral statistics in this scenario, which embodies exact self-similarity of the spectral correlations along the complete scrambling dynamics. We establish that the scaling predictions are matched by a privileged stochastic process and serve as bounds for other dynamical scrambling scenarios, allowing one to quantify inefficient or incomplete scrambling on all time scales.
\end{abstract}
\section{Introduction}
A central theme in the study of complex quantum matter is to establish universal characteristics that transfer between systems and application domains. This theme unifies different areas of physics, spanning from its historic origins in nuclear physics \cite{Wishart1928ThePopulation, Wigner1955CharacteristicDimensions, Wigner1958OnMatrices,
Dyson1962AMatrix, Dyson1962StatisticalI, Mehta1967RandomLevels}, to disordered and wave-chaotic electronic and photonic systems \cite{Beenakker1997RandomTransport, GUHR1998189, Akkermans2007MesoscopicPhotons, Stockmann1999QuantumChaos, Haake2018QuantumChaos}, to isolated interacting models that display many-body eigenstate thermalization \cite{Deutsch1991QuantumSystem, Srednicki1994ChaosThermalization, Nandkishore2015Many-bodyMechanics, DAlessio2016FromThermodynamics, Borgonovi2016QuantumParticles, Abanin2019iColloquium/iEntanglement}, and also provides insights into the black hole information paradox \cite{Page1993AverageSubsystem, Page1993InformationRadiation, Hayden2007BlackSubsystems, Lashkari2013TowardsConjecture, Shenker2014BlackEffect, Maldacena2016AChaos}.
Across all these settings, a universal endpoint of the dynamics can be defined in terms of random-matrix theory (RMT) and systems that approach this endpoint are described as being ergodic. Universal random-matrix behavior also sets benchmarks for systematic deviations reflecting the specific structure of a system, such as those observed in the interplay of short-range interactions, disorder, and conservation laws \cite{Bohigas1984CharacterizationLaws, Brezin1997SpectralTheory, Stockmann1999QuantumChaos, Haake2018QuantumChaos}. 
Even for systems that establish ergodicity over time, the approach to this endpoint itself---the dynamical process known as scrambling---is system-dependent \cite{Hosur2016ChaosChannels}. This leads to the emergence of characteristic time and energy scales imprinted onto the dynamics in addition to the concept of maximally chaotic systems, which are posited to display universal characteristics already for short times \cite{YasuhiroSekino2008FastScramblers, Maldacena2016AChaos}.
For maximal chaos, the conjectured state outside the horizon of a black hole serves as an important motivation \cite{YasuhiroSekino2008FastScramblers, Lashkari2013TowardsConjecture, Shenker2014BlackEffect, Maldacena2016AChaos,Sachdev1993GaplessMagnet, KitaevAlexei1, Maldacena2016RemarksModel, Hayden2007BlackSubsystems, Hosur2016ChaosChannels, Garcia-Garcia2016SpectralModel, Cotler2017BlackMatrices, Saad2019AGravity}.

Mathematically, these emergent universal features can be expressed in terms of chaos bounds that govern, for instance, the behavior of out-of-time-ordered correlators (OTOCs) \cite{Maldacena2016AChaos, Cotler2017BlackMatrices}. 
Since the latter primarily capture operator growth in initially disjoint supports, they are generally associated to operator scrambling and, albeit typically conflated with, have recently been shown to be insufficient for diagnosing quantum chaos \cite{Dowling2023ScramblingChaos}.   
Open system dynamics involve continuous information exchange with the environment, making scrambling more akin to decoherence than classical chaos \cite{Zanardi2021InformationSystems, Tripathy2024QuantumLattices}. As such, studying the OTOC in isolation is not enough to fully discriminate between unitary evolution inducing scrambling and external noise responsible for decoherence in experiments \cite{Touil2021InformationEntropy,  Touil2020QuantumInformation}, motivating the study of other quantities that include the Krylov complexity \cite{Hornedal2022UltimateComplexity, Hashimoto2023, Carolan2024OperatorSystems}.
On the other hand, from a spectral perspective, any scrambling dynamics are subject to strict unitarity constraints, which enforce a duality between times shorter and longer than the Heisenberg time
\cite{MVBerry1990AChaos, Berry1992ADeterminants, EBBogomolny1992SemiclassicalSystems, Keating1992PeriodicChaos, Winer2022HydrodynamicFactor, Winer2023ReappearanceFactor}. 
This spectral perspective implies that maximally ergodic long-time behavior provides universal constraints on the short-time scrambling dynamics that are directly imprinted onto the spectral statistics.

In this work, we utilize this spectral perspective to develop a predictive single-parameter scaling theory for the efficient scrambling dynamics of maximally chaotic systems and apply this theory to obtain analytical benchmarks for their behavior over all time scales, formalized as spectral chaos bounds. 
This uncovers universality in the language of a general framework that relates all statistical details to a single intrinsic parameter \cite{Abrahams1979ScalingDimensions}.
Our scaling assumption is simple---we equate the only two invariants of the dynamics under the assumption that the Hilbert space has no further structure, i.e., that the dynamics are invariant under unitary basis changes. This ansatz integrates into a single-parameter version of a specific random-matrix ensemble, the Poisson kernel, which has been widely studied in static settings \cite{Hua1963HarmonicDomains, Mello1985InformationProblems, Beenakker1997RandomTransport, Szyniszewski2020UniversalityMeasurements} but in our scaling theory acquires a dynamical interpretation, where it embodies exact self-similarity of the spectral correlations along the complete scrambling dynamics.

We furthermore apply this scaling framework to evaluate the scrambling dynamics in general unitarily invariant processes. Among these, we establish that the scaling predictions are faithfully replicated in a second central object of random-matrix theory, the paradigmatic Dyson Brownian motion process \cite{Dyson1962AMatrix}, which similarly shifts its role from being a tool to study the stationary ergodic endpoint to become a dynamical process in its own right {\cite{Ito1950BrownianGroup, Yosida1952OnSpace, Hunt1956Semi-groupsGroups, Katori2016BesselModel, Erdos2017ATheory, Schomerus2022NoisySystems, Forrester2024Dip-ramp-plateauUN}. 
Utilizing exact analytical expressions for the scaling parameter, density of states, and spectral correlation functions, supplemented by numerical results, we find that the spectral data from this process agrees with the scaling theory and recovers its key features, such as the self-similarity of correlations along the flow.
Moreover, a recent work \cite{Kalsi2024HierarchicalUniversal} studies the emergent universal correlations of stochastic models directly connected to our work, particularly Dyson's Brownian motion generated within the three Wigner-Dyson symmetry classes that are defined by the presence and nature of time-reversal symmetry, and shows that findings from these models further extend to the Brownian Sachdev-Ye-Kitaev model \cite{Saad2019AGravity, Sunderhauf2019QuantumInformation}. 
Therefore, from a wider perspective, we anticipate that the results reported here apply to a much larger class of systems. 
On the other hand, we also contrast this behavior with another unitarily invariant stochastic process that displays clear deviations from the scaling predictions, thereby illuminating the role of these predictions as spectral chaos bounds. As our theory manifestly preserves all unitarity constraints, it emphasizes the role of functional relations linking the short- and long-time dynamics, both in the universal regime as well as deviations away from it, from which we can draw broader conclusions about the approach to ergodicity in complex quantum matter.

This paper is organized as follows. 
We begin by reviewing the connection between scrambling and chaos bounds in Section~\ref{sec:background} and contrast the out-of-time-ordered correlators typically used to formulate these bounds with the
spectral form factor, which is the focal quantity in our formulation in terms of spectral statistics.
In Section~\ref{sec:scaling}, we develop the scaling theory and derive analytical forms of the spectral form factor for optimally efficient scrambling. 
In Section~\ref{sec:discussion}, we utilize the scaling framework to characterize specific processes and demonstrate that scaling expressions serve as sensitive chaos bounds. 
In Section~\ref{sec:otoc}, we further extend our considerations of scrambling dynamics to calculate the out-of-time-ordered correlator in the scaling theory explicitly. 
Finally, we draw our conclusions in Section~\ref{sec:conclusions}. 

\section{Background and objective} \label{sec:background}

\subsection{Characterizing scrambling}
While complex quantum many-body dynamics vary greatly for different quantum systems, it is conjectured that there exists an upper bound on how quickly such systems can disperse, or \textit{scramble}, local information into nonlocal degrees of freedom \cite{YasuhiroSekino2008FastScramblers, Maldacena2016AChaos, Cotler2017BlackMatrices,Xu2024ScramblingSystems}. 
A principal diagnostic tool to quantify these scrambling dynamics is the out-of-time-ordered correlator (OTOC) between two arbitrary observables, say $V(0)$ and $W(t)$, at time separation $t$ \cite{Maldacena2016AChaos, Cotler2017BlackMatrices}, such that the OTOC probes the perturbative effect by $V$ on later measurements of $W$, and vice versa. 
The Maldacena-Shenker-Stanford bound conjectures that this correlator develops at most exponentially \cite{Maldacena2016AChaos} in time, and never faster than a universal Lyapunov exponent, setting a bound that is independent of the details of the system. 
This \textit{chaos bound} is saturated by \textit{fast scramblers}, thought to encompass systems such as black holes and the Sachdev-Ye-Kitaev (SYK) model at low temperature \cite{Shenker2014BlackEffect, Sachdev1993GaplessMagnet, KitaevAlexei1}. 
After the scrambling time, this initial exponential growth of the OTOC for chaotic systems \cite{Rozenbaum2017LyapunovSystem, Jalabert2018SemiclassicalSystems, Lakshminarayan2019Out-of-time-orderedMatrices} settles into saturation oscillations, whose amplitude is suppressed in the chaotic limit \cite{Li2017MeasuringSimulator, Hummel2019ReversibleCriticality, Fortes2019GaugingCorrelators, Fortes2020SignaturesChains}, and generally displays nonuniversal behavior \cite{Dora2017Out-of-Time-OrderedLiquids, Ali2020ChaosMechanics, Xu2020DoesChaos, Hashimoto2020ExponentialOscillator, Morita2022ExtractingMechanics}.
\textit{Slow scramblers} fail to saturate this bound on all time scales, only attaining, e.g., logarithmical growth as in many-body localized systems \cite{Huang2017OutoftimeorderedSystems, Fan2017Out-of-time-orderLocalization, Swingle2017SlowSystems}, linear short-time growth as in weakly chaotic systems \cite{Kukuljan2017WeakChaos}, or quadratic growth as in Luttinger liquids \cite{Dora2017Out-of-Time-OrderedLiquids}. 

This interplay between universal and nonuniversal features replicates a common theme known from the study of spectral statistics. In this spectral setting, the key quantity to capture both the universal and system-specific aspects of the dynamics over all time scales is the spectral form factor (SFF) \cite{Haake2018QuantumChaos, Stockmann1999QuantumChaos, Prange1997TheSelf-Averaging, Brezin1997SpectralTheory, Berry1985SemiclassicalRigidity, Sieber2001CorrelationsStatistics, Muller2004SemiclassicalChaos, Muller2005Periodic-orbitChaos, Richter2022SemiclassicalChaos, Garcia-Garcia2016SpectralModel, Cotler2017BlackMatrices, Saad2019AGravity, Chan2018SolutionChaos, Kos2018Many-BodyTheory, Chan2018SpectralSystems, Bertini2018ExactChaos, Friedman2019SpectralCharge, Bouverot-Dupuis2024RandomTimes,  Yang2020QuantumHamiltonian, Joshi2022ProbingSimulators, Dag2023Many-bodyAtoms, Dong2024MeasuringProcessors, Das2024ProposalDetection}. 
While originally defined as the Fourier transform of the two-point level correlation function, for systems with a finite Hilbert-space dimension $N$, the SFF can be defined directly in terms of the unitary time-evolution operator $U(t)$,
\begin{equation} \label{eq:SFF}
    K(t) \equiv \overline{|\mathrm{tr}\, U(t)|^2 }. 
\end{equation}
Since the SFF is not self-averaging \cite{Prange1997TheSelf-Averaging}, the overline $\overline{[...]}$ denotes a suitable average, which may, for instance, take the form of an additional ensemble average over statistically similar systems as is the case in this work, or, in some settings, including high-energy physics, a partition sum \cite{Cotler2017BlackMatrices}. 
Employing such averaging within random-matrix theory, namely averaging over an ensemble of random matrices, particularly simple results arise for the case of stroboscopic dynamics of period $T$, where $U(nT)=U^n(T)$, and
\begin{equation} \label{eq:stroboscopicSFF}
    K_n\equiv \overline{|\, \mathrm{tr} \, U^n(T)|^2}.
\end{equation}
In the circular unitary ensemble (CUE), one then obtains
\begin{equation}
\label{eq:sffcue}
    K_n=N^2\delta_{n0}+\mathrm{min}\,(n,N), 
\end{equation} 
which dips abruptly from $K_0=N^2$ to $K_1=1$, and then ramps up steadily from unity to $N$ over the course of the effective stroboscopic Heisenberg time $n=N$. However, this result does not capture the details of the short-time scrambling dynamics itself.
For complex quantum systems, $K(t)$ generally displays a dip over the scrambling regime, taking the value of unity at the minimum, followed by a ramp up to $K(t)\sim N$ until the Heisenberg time, at which the discreteness of the level spectrum becomes resolved, which then is followed by a plateau. System-specific signatures can persist well into the ramp, while fast scramblers are expected to display universality already during the dip \cite{YasuhiroSekino2008FastScramblers, Gharibyan2018OnsetSystems, Chan2018SpectralSystems, Chan2022Many-bodyInvariance}. When expressed in terms of the energy levels, the SFF captures their correlations at all scales, including level repulsion and spectral rigidity, and thus directly gives information about basic system properties such as integrability or time-reversal symmetry \cite{Haake2018QuantumChaos, Stockmann1999QuantumChaos, Suntajs2020QuantumLocalization}. 

Our general objective is to combine these themes and establish universal benchmarks of efficient complex quantum dynamics in terms of spectral information. This is paired with the conjecture that for efficient systems, this spectral information again recovers a strong degree of universality.

\subsection{Spectral form factor and unitarily invariant processes}

We will phrase our results in the context of
stochastic processes
\begin{equation} \label{eq:stochastic}
    U(t)\to U(t+dt)=u(t;dt)U(t),
\end{equation} in which the unitary matrix $U(t)$ generating the dynamics is updated incrementally by unitary matrices $u(t;dt)\simeq\openone$ over a small time step $dt$. 
These processes describe a large class of dynamics, among which we want to identify and characterize those that facilitate maximally efficient chaotic dynamics. 

We prepare this discussion by adapting the spectral form factor to this setting. 
At any point of time $T$, this dynamical evolution can be equipped with a definite Heisenberg time by considering the instantaneous SFF \eqref{eq:stroboscopicSFF}.
While this can again be interpreted as a Floquet dynamics in which, from a certain time $T$, the evolution generated by the unitary time-evolution operator $U(T)$ is repeated periodically, the instantaneous SFF is well-defined independent of this interpretation and analyzes the moments of $U(T)$, thus probing the dynamics over all energy and time scales \cite{Kos2021ChaosDriving, Forrester2024Dip-ramp-plateauUN, Kalsi2024HierarchicalUniversal}. 
In the context of this work, it facilitates detailed analysis by giving us two time scales---the time $T$ for the evolution along the scrambling dynamics, and the time $nT$ resolving the spectral statistics established up to this point. 

Our approach will allow us to relate these spectral correlators through the flow of a single scaling parameter that we introduce in the next section.
To establish some early intuition for what we are aiming at, let us apply the definition \eqref{eq:stroboscopicSFF} to any dynamical evolution \eqref{eq:stochastic} induced by ensembles of generators that are invariant under rotations $u(t;dt)\to W^\dagger u(t;dt) W$. Evaluating the average over $W$ in the CUE, the first-order SFF incrementally updates as
\begin{align} \label{eq:Invariance}
    &K_1(t+dt)\\
    \quad&=K_1(t)+\frac{N^2-\overline{|\,\mathrm{tr}\, u(t;dt)|^2}}{N^2-1}\left(1-K_1(t)\right),
    \nonumber
\end{align}
where the overline here denotes averaging over the specific ensemble generating stochastic dynamics, resulting in an exponential decay to unity with the decay constant
\begin{equation} \label{eq:gamma1}
\gamma_1=\lim_{dt\to 0}dt^{-1}(N^2-\overline{|\,\mathrm{tr}\, u(t;dt)|^2})/(N^2-1).
\end{equation}
Within this class of unitarily invariant dynamics, distinct processes, each characterized by some $\overline{|\,\mathrm{tr}\, u(t;dt)|^2}$, are therefore characterized by different rates $\gamma_1$, which discriminate how efficiently they approach ergodic random-matrix behavior. Moreover, it follows directly from the statistical definitions that the corresponding decay rates of the higher-order SFFs are constrained as $\gamma_n \leq n \gamma_1$.

Our specific objective is to turn such relations into spectral chaos bounds for the maximally efficient scrambling scenario, which abstracts away the arbitrary overall time scale of the dynamics. For this, we set out to formulate such bounds in a stricter fashion in terms of a single scaling parameter.

\section{Scaling theory of the spectral form factor} \label{sec:scaling}

\subsection{Scaling ansatz}
For maximally chaotic scrambling, we expect the dynamics to have minimal constraints. On the level of the incremental time evolution \eqref{eq:stochastic}, this would include representative processes for which the generators $u(t;dt)$ are invariant under unitary basis changes. We now observe that on the level of the  resulting  dynamics $U(t)$ at finite times, there are two fundamental anti-Hermitian invariants, $U^\dagger \frac{dU}{dt}$ and $U - U^\dagger$, that are also invariant under unitary basis changes. Of these, the former extracts the generator of the time evolution, while the latter characterizes the departure from the initial conditions.
Constraining the description of these processes in terms of such invariants is in the spirit of the seminal scaling theory of localization that relates the fluctuations in the conductance to the value of the conductance itself \cite{Abrahams1979ScalingDimensions}.
For maximally chaotic scrambling, where we expect the dynamics to have no further constraints, we therefore propose the scaling assumption
\begin{equation} \label{eq:Scaling}
  U - U^\dagger = g(t) U^\dagger \frac{dU}{dt}, 
\end{equation}
which equates these invariants in the ensemble sense up to a time-dependent factor $g(t)$.

This scaling ansatz can be integrated by introducing the parameter
\begin{equation}
a(t)=\mathrm{tanh}\left[\mathrm{atanh}\,(a_0)-\int_{t_0}^t (1/g(t')) dt'\right],
\end{equation}
which follows from setting
$g(t)d/dt=(a^2-1)d/da$.
The scaling ansatz \eqref{eq:Scaling} is 
then solved by the parameterized ensemble
\begin{equation} \label{eq:PoissonKernel}
    U = (a\openone+V)(\openone+ aV)^{-1}.
\end{equation}
The originally assumed basis invariance is respected when $V$ is uniformly distributed in the unitary group of degree $N$, hence, taken from the CUE. 

The resulting ensemble is a single-parameter incarnation of the Poisson kernel, a matrix ensemble that previously appeared in stationary scattering settings subject to some constraint \cite{Hua1963HarmonicDomains, Mello1985InformationProblems, Beenakker1997RandomTransport, Szyniszewski2020UniversalityMeasurements}, where its functional form is tied to a multiple-scattering expansion \cite{Brouwer1995GeneralizedLeads}.
Here, we encounter it instead in the context of dynamics generated by a multiplicative composition law, where the dynamical flow of the scaling parameter $a(t)$ will be of central importance.

\begin{figure}[t]
    \centering
    \includegraphics[width=\columnwidth]{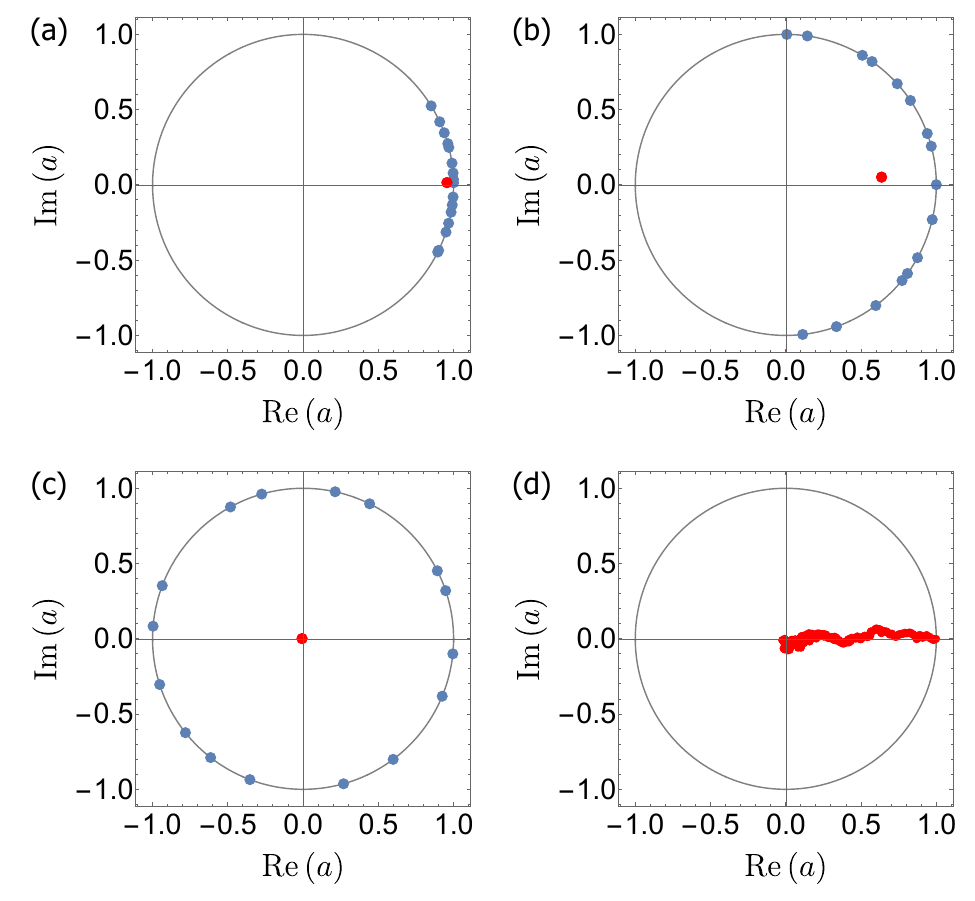}
    \caption{Interpretation of the spectral scaling parameter $a$ as the center of mass (red) of the eigenvalue distribution (blue), illustrated for a single time evolution generated by the multiplication of random unitary matrices of the form \eqref {eq:uWiener} ($N=16$, $dt=0.01$). Panels (a-c) show snapshots after $10$, $100$, and $1000$ time steps, while panel (d) shows the complete center-of-mass trajectory over $1000$ time steps.
    }
    \label{fig:cloud}
\end{figure}

\subsection{Interpretation of the scaling parameter} 
Our scaling assumption reduces the matrix-generated scrambling dynamics to a single dynamical scaling parameter $a$. 
In terms of this parameter, the ensemble \eqref{eq:PoissonKernel} interpolates between action by the identity ($U=\openone)$ at $a=1$ and the random unitary matrix $U=V$ at $a=0$, i.e., the static ergodic endpoint defined by the CUE.
For intermediate time, we can equate $a=N^{-1} \overline{\mathrm{tr}\,U}$, with the overline representing an ensemble average of the dynamics, to characterize the motion of the center of mass of the eigenvalues $\lambda_n\equiv\exp(i\phi_n)$, capturing their expansion on the unit circle as ergodicity is established. This center-of-mass motion is illustrated in an individual realization $U(t)$ in Fig.~\ref{fig:cloud}. The cloud of eigenvalues, initially centered at unity, begins to disperse around the unit circle such that $N^{-1} \mathrm{tr}\,U$ performs a stochastic trajectory towards the origin---the RMT result---where the center of mass of the eigenvalues is zero.

This motion is tied to a specific scaling of the density of states, which we will have to take into account in the application of the theory.
The mean density of eigenvalues $\lambda_l\equiv\exp(i\phi_l)$ of the 
unitary time-evolution operator $U$ can be obtained by first expressing it in terms of the moments $A_n=N^{-1}\overline{\mathrm{tr}\,U^n}$, whereby
\begin{equation}
\label{eq:dos}
 \rho(\phi)=\frac{1}{2\pi}\left(1+2\sum_{n=1}^\infty A_n \cos{(n \phi)} \right).
\end{equation}
In the scaling theory, $A_n = a^n$ follows directly by expanding Eq.~\eqref{eq:PoissonKernel} into a geometric series, and using the CUE average $\overline{V^m} =\delta_{0m}\openone$. 
The first term
\begin{equation}
A_1=a=N^{-1}\overline{\mathrm{tr}\,U}
\label{eq:aapp}
\end{equation}
recovers the interpretation of the scaling parameter as the center of mass of the eigenvalue cloud.
Summation of the series \eqref{eq:dos} then delivers the scaling mean density of states
\begin{equation}
\rho(\phi)= \frac{1}{2\pi} \frac{1-a^2}{1+a^2-2a\cos{\phi}}, 
\label{eq:scalingdos}
\end{equation}
whose flow corresponds to the expansion of the eigenvalues. 

The scaling ensemble \eqref{eq:PoissonKernel} endows this expanding eigenvalue cloud with intrinsically universal spectral statistics, induced via the matrix-valued Mobius transformation
\begin{equation}
U'=\left(\frac{a'-a}{1- aa'}\openone+U\right) \left(\openone+\frac{a'-a}{1-aa'} U\right)^{-1}
\label{eq:mobius}
\end{equation}
between the ensembles with parameters $a$ to $a'$.
At the same time, the scaling parameter defines a specific unfolding procedure of the eigenvalues. For any pair of unitary matrices $U$, $V$ related by \eqref{eq:PoissonKernel}, the eigenvalues $\mu_l=\exp(i \psi_l)$ of $V$ determine the eigenvalues 
\begin{equation}
     \label{eq:lambdamapping}
     \lambda_l=(a+\mu_l)/(1+a \mu_l) = \exp(i \phi_l)
\end{equation}
of $U$. When $V$ is sampled from the CUE, $\lambda_l$ are eigenvalues distributed as in the Poisson kernel with the scaling mean density of states \eqref{eq:scalingdos}. This transformation can be inverted to translate the eigenvalues $\lambda_l$ into uniformly distributed eigenvalues 
    \begin{equation}
     \label{eq:mumap}
  \mu_l =(a-\lambda_l)/(a \lambda_l-1), 
    \end{equation}
constituting an unfolding procedure in which eigenvalues are unfolded to a uniform (CUE) distribution, at any instant of time $t\neq0$ in the evolution. 

\subsection{Derivation of the spectral form factor scaling predictions} \label{sec:sffderive}

Within the scaling ensemble \eqref{eq:PoissonKernel}, we can analytically analyze the SFF by expressing it in terms of the eigenvalues $\lambda_l$ of $U$, 
\begin{equation}
    K_n =\overline{|\, \mathrm{tr} \, U^n|^2} =\sum_{lm} \lambda_l^n\lambda_m^{*n}. 
\end{equation}
To this end, we use the transformation \eqref{eq:lambdamapping} to recast the SFF
in each realization $U$ explicitly in terms of the eigenvalues
$\mu_l=e^{i\psi_l}$ of the corresponding CUE matrix $V$,
\begin{equation}
    |\mathrm{tr} \, U^n|^2 = \sum_{lm} \left( \frac{a+e^{i\psi_l}}{1+ae^{i\psi_l}} \frac{a+e^{-i\psi_m}}{1+ae^{-i\psi_m}}\right)^n.
\end{equation}
The joint distribution
\begin{equation}
    P_\mathrm{CUE}(\{\psi\}) \propto \prod_{l<m} \left|e^{i\psi_l} - e^{i\psi_m}\right|^2 
    = \mathrm{det}\left(\sum_l \mu_l^{p-q}\right)
\end{equation}
of these eigenvalues can be written as a product of two Vandermonde determinants, where the indices $q,p=1,2,...,N$ label the rows and columns of the resulting determinant.
While this does not factorize in terms of the eigenvalues $\mu_l$, Ref.~\cite{Haake1996SecularMatrices} establishes that the average of any completely symmetric function $f(\{\psi_l\})$ simplifies to
\begin{equation} \label{eq:proof}
    \overline{ f(\{\psi_l\})} =  \left(\prod_r\int_0^{2\pi}  \frac{d\psi_r}{2\pi}\right) f(\{\psi_l\}) \mathrm{det}\left( \mu_p^{p-q}\right).
\end{equation}
Applied to the SFF, we then obtain 
\begin{align} \label{eq:StroboscopicFFapp}
    {K_n}
     =& \left( \prod_r \int_0^{2\pi} \frac{d\psi_r}{2\pi} \right) \mathrm{det}(e^{i(p-q)\psi_q})
     \nonumber\\
&\times \sum_{lm} \left(\frac{a+e^{i\psi_l}}{1+ae^{i\psi_l}}\right)^n
\left(\frac{a+e^{-i\psi_m}}{1+ae^{-i\psi_m}}\right)^n
.
\end{align}
In this expression, the integrals over $\psi_q$ can be performed independently of one another and pulled into the $q$th rows of the matrix in the determinant.
Each of the diagonal terms $l = m$ in the sum gives a contribution of 1, while the off-diagonal terms $l \neq m$ give contributions
\begin{equation} \label{eq:off-diag}
     \mathrm{det} \begin{pmatrix}
     a^n & c_{m-l,n} \\
     c_{m-l,n} & a^n
     \end{pmatrix} = a^{2n} - c_{m-l,n}^2,
\end{equation}
where the integrals
\begin{equation} \label{eq:coeff}
    c_{q,n} =  \int_0^{2\pi} \frac{d\psi}{2\pi} \left( \frac{a+e^{-i\psi}}{1+ae^{-i\psi}}\right)^n e^{i\psi q}
\end{equation}
correspond to the coefficient $\propto v^q$ in the expansion of $(a+v)^n/(1+av)^n$ in powers of $v$ and hence are finite only for $m>l$. 

Altogether, we arrive at
\begin{align}
    {K_n} &= N + {N(N-1)} a^{2n} - \sum_{q=1}^{N}(N-q)c_{q,n}^2,
\nonumber    \\
    c_{q,n}&=\frac{1}{q!}\frac{d^q}{dv^q} \left.\frac{ (a+v)^n}{(1+av)^n}\right|_{v=0}.
\label{eq:StroboscopicFFResult}
\end{align}
The first term is due to each diagonal $l=m$ term contributing 1, summing to $N$ overall. The second term is due to $(N^2 - N)$ distinct off-diagonal terms with $l\neq m$, each contributing $a^{2n}$ from Eq.~\eqref{eq:off-diag}. The third then collects the contributions $-c_{q,n}^2$ from this equation over all possible combinations of $m$ and $l$ with fixed $q=m-l \geq 1$ permitted by the matrix of dimension $N$. Equation~\eqref{eq:StroboscopicFFResult} recovers the standard CUE result for $a=0$, where $c_{q,n}=\delta_{qn}$ such that $K_n$ take the form given in Eq.~\eqref{eq:sffcue}.

Equation \eqref{eq:StroboscopicFFResult} is our main result within the scaling theory. It expresses all orders of the SFF in terms of a single parameter, which has its independent interpretation as characterizing the expansion of the eigenvalue cloud. Next, we will describe how this result can be utilized as a benchmark to analyze specific dynamical processes.

\section{Applications} 
\label{sec:discussion}
To establish how the scaling forms \eqref{eq:StroboscopicFFResult} of the SFF provide benchmarks for maximally ergodic dynamics, we describe how they can be used to distinguish the effects of efficient but incomplete scrambling and inherently inefficient scrambling. We develop these features both within the scaling framework itself, as well as in the context of two specific stochastic processes. 

\subsection{Efficient but incomplete scrambling }

First, we describe how the effects of incomplete scrambling up to a given time $T$ are captured within the scaling approach.
As mentioned above, the scaling forms \eqref{eq:StroboscopicFFResult} of the instantaneous SFF reduce to the CUE result \eqref{eq:sffcue} for $a=0$, which defines the endpoint of the scrambling flow. The SFF then falls from $K_0=N^2$ to $K_1=1$ before ramping up linearly to $K_N=N$ at the stroboscopic Heisenberg time $N$, after which it plateaus. Figure~\ref{fig:Stroboscopic}(a) contrasts this behavior of the instantaneous SFF \eqref{eq:StroboscopicFFResult} with the scenarios for finite values of $a$. Tuning the scaling parameter $a$ away from 0---stopping the dynamics at time $T$ short of maximally ergodic behavior---results in curves that initially continue to dip, which then take a longer time to recover the plateau. Therefore, incomplete scrambling dynamics at short times are translated into a long-time signal in the form of a modified ramp, demonstrating the consequences of not having established fully ergodic dynamics for the remainder of the time evolution. The time over which the curves continue to dip defines an effective ergodic time, while the time it takes them to ramp up to the plateau defines an effective Heisenberg time. Crucially, within the scaling theory, these two time scales are directly linked via the scaling parameter $a$.

This link is emphasized by the scaling relation between these results.
The transformation \eqref{eq:mobius} directly transfers into self-similar correlations of the eigenvalues $\lambda_l$ along the flow.
Unfolding the spectrum to a uniform density with Heisenberg time $N$ according to \eqref{eq:mumap} collapses the SFF identically onto the RMT result, as illustrated in Fig.~\ref{fig:Stroboscopic}(b). Within the scaling ensemble, this collapse is exact, underlining both its scale invariance and single-parameter nature.

\begin{figure}[t]
    \includegraphics[width=\columnwidth]{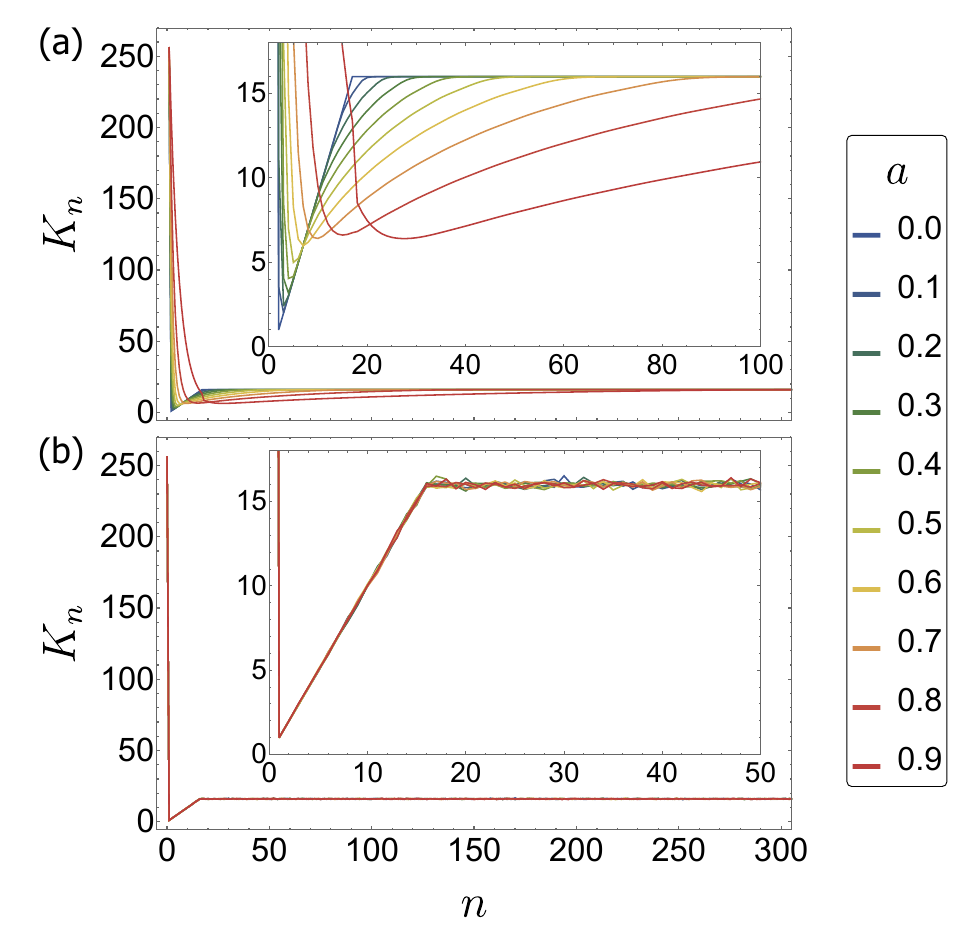} 
    \caption{
    (a) Scaling predictions of the instantaneous form factor $K_n$ for maximally efficient scrambling, Eq.~\eqref{eq:StroboscopicFFResult}, for $N=16$, where $a$ describes different points along the scrambling flow. All curves display the paradigmatic dip-ramp-plateau shape. For $a=0$, scrambling is complete, and the curves follow the RMT predictions of an ergodic system. Finite values of $a$ describe earlier times along scrambling dynamics, resulting in effective ergodic and Heisenberg times that are linked by the scaling parameter $a$.
    (b) Numerical sampling of the ensemble ($10^4$ realizations) confirms that the points along the scrambling flow are linked by the transformation \eqref{eq:mobius}, which implies self-similar statistics and the exact collapse onto the RMT result after unfolding the spectrum according to \eqref{eq:mumap}, corresponding to setting $a'=0$ in Eq.~\eqref{eq:mobius}. 
    }
    \label{fig:Stroboscopic}
\end{figure}

\subsection{Dyson's Brownian motion: a manifestation of efficient scrambling}
We next turn to question whether this single-parameter behavior within the ensemble can be replicated in a suitable unitary time evolution. Which dynamical process, if any, recovers the statistics of the scaling ensemble \eqref{eq:PoissonKernel}, parametrized by a single suitable time-dependent scaling parameter $a$? We argue that the answer lies in another paradigm of RMT, Dyson's Brownian motion (DBM).

DBM emerges as a natural candidate for fast scrambling in the context of quantum circuit models. These come in two main variants: random Haar circuits (e.g., \cite{Li2018QuantumTransition,
Nahum2018OperatorCircuits, Skinner2019Measurement-InducedEntanglement, Li2019Measurement-drivenCircuits}), built out of fully ergodic gates from RMT, and Brownian circuits \cite{Brown2012ScramblingCircuits, Lashkari2013TowardsConjecture, Shenker2015StringyScrambling, Gharibyan2018OnsetSystems, Chen2019QuantumSystems, Zhou2019OperatorCircuit, Xu2019LocalityScrambling, Zhou2020OperatorSystems}, 
built from gates with randomly chosen Hamiltonians $H(t)$ over small time steps $dt$. Our scaling approach interpolates between both types of models for one of these gates, and so does the Brownian process applied for a finite time.
This coincides with the DBM process, where the unitary time-evolution operator $U(t)$ performs a random walk in the unitary group, sampling it uniformly according to the Haar measure. 

Originally, this process was introduced to facilitate RMT calculations \cite{Dyson1962AMatrix}, 
and has since served as a central tool in celebrated proofs of universality over a broad class of RMT models \cite{Tao2011RandomStatistics, Tao2012TopicsTheory}. 
Here, we consider it as a genuinely dynamical model with a specific initial condition, $U(0)=\openone$.
This is implemented by generating incremental unitary time steps
\begin{equation} \label{eq:uWiener}
    u(t;dt) = \left(\openone- \frac{iH(t)}{2}\sqrt{dt}\right)\left(\openone+ \frac{iH(t)}{2}\sqrt{dt}\right)^{-1}
\end{equation}
with an instantaneous Hamiltonian $H(t)$ from the Gaussian unitary ensemble (GUE), given by normally distributed matrix elements satisfying $\overline{H(t)_{lm}}=0,\, \overline{H(t)_{kl}H(t)_{mn}}=N^{-1}\delta_{kn}\delta_{lm}$.

Within this process, we find that the scaling parameter $a$ decays exponentially from unity to zero, $a(t) = e^{-t/2}$, corresponding to a dimensionless decay rate $\gamma_0=1/2$. 
This is accompanied by an exponential decay $K_1(t) = (N^2-1)e^{-t}+1$
of the first-order SFF, describing the dip to unity with a decay rate $\gamma_1=2\gamma_0$, which agrees with the scaling prediction \eqref{eq:StroboscopicFFResult} up to corrections $O(N^{-2})$.
The key question is whether this process recovers the complete spectral statistics encoded in the scaling forms \eqref{eq:StroboscopicFFResult} of $K_n$, displaying self-similar spectral correlations up to standard unfolding, as in the scaling theory itself.

This is analyzed in Fig.~\ref{fig:stochastic}. The top panel shows the SFF after unfolding the DBM spectrum to the scaling mean density of states \eqref{eq:scalingdos} (see Appendix~\ref{app:b}). We observe that this agrees with the scaling prediction \eqref{eq:StroboscopicFFResult} up to statistical fluctuations over the whole range of the scaling parameter, hence, over the complete scrambling dynamics. Furthermore, upon fully unfolding the spectrum to a uniform mean density in the bottom panel, we find perfect collapse of all data onto straight lines $K_n=n$, which establishes agreement with the scaling theory down to the level of self-similarity under the flow, again on all scales of $a$.

As we show next, this tight agreement---including the higher orders of the SFF---is a nontrivial statement about the DBM process, marking it out as a privileged model of fast scrambling among a wider class of dynamical models.

\begin{figure}[t]
    \includegraphics[width=\columnwidth]{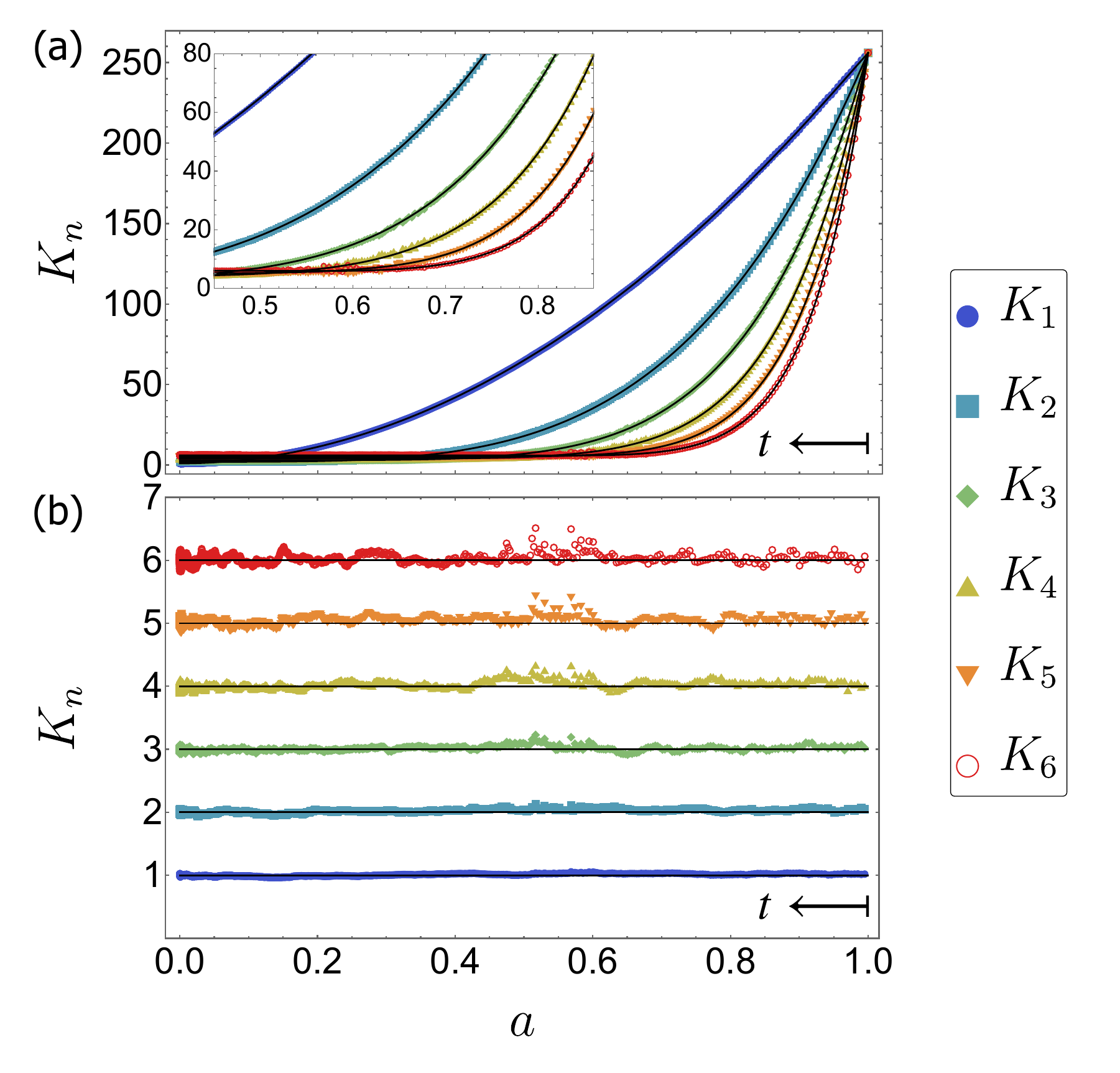} 
    \caption{Scaling analysis of scrambling in the DBM process, generated by Eq.~\eqref{eq:uWiener} ($N=16$, $dt=0.01$, $10^4$ realizations).
   (a) Spectral form factors $K_n$ after unfolding the spectrum to the scaling mean density of states \eqref{eq:scalingdos} at $a(t)=e^{-t/2}$, as a function of $a$. There is agreement within statistical uncertainty with the analytical scaling predictions \eqref{eq:StroboscopicFFResult} (black curves). 
   (b) Further unfolding the spectrum to uniform density collapses it onto the RMT prediction $K_n=n$, verifying that DBM generates self-similar spectral statistics along the complete scrambling dynamics.}
    \label{fig:stochastic}
\end{figure}

\subsection{Deviations from efficient scrambling: formulation via chaos bounds}
Our scaling ansatz \eqref{eq:Scaling} equates two unitarily invariant generators. 
This unitary invariance is also obeyed in DBM and is the sole constraint in the derivation of the exponential decay of the first-order SFF \eqref{eq:Invariance} to unity with decay rate \eqref{eq:gamma1}. 
Studying this decay of the SFF in isolation, one could be led to believe that all systems within this more general class of unitarily invariant processes exhibit maximally chaotic scrambling. A first hint that this may not be the case is given by the decay of the scaling parameter itself, where we again observe an exponential decay, but with a decay rate (see Appendix~\ref{app:c})
\begin{equation} \label{eq:gamma0}
\gamma_0=\lim_{dt\to 0}(dt\,N)^{-1}(N-\overline{\mathrm{tr}\, u(t;dt)})
\end{equation}
that is not universally linked to $\gamma_1$ given by Eq.~\eqref{eq:gamma1}. Instead, the mathematical definitions of these quantities enforce the relation $\gamma_1\leq 2 \gamma_0$, again up to corrections $O(N^{-2})$.
The scaling forms \eqref{eq:StroboscopicFFResult} satisfy this constraint tightly, and also its extension to the decay rates of the higher-order SFFs, $\gamma_n \leq 2n \gamma_0$. 
We can therefore view these scaling forms as lower bounds that are approached only for maximally scrambling dynamics, as represented, e.g., by DBM. 
Within the class of unitarily invariant processes, we can view these bounds for the decay of the $n$th-order SFF as the analog of the chaos bounds on OTOCs, here obtained directly from the spectral statistics. Moreover, by further adapting our scaling analysis of the DBM process, we can take one further step and directly utilize the scaling forms \eqref{eq:StroboscopicFFResult} of the SFF as bounds that are expressed in terms of the scaling parameter $a$.

We exemplify this by modifying DBM, which mathematically corresponds to a Wiener process, into a Cauchy process, obtained from generators
\begin{equation} \label{eq:uCauchy}
u(t;dt)=\left(\sqrt{1-dt}\openone+V\right)\left(\sqrt{1-dt}V+\openone\right)^{-1}
\end{equation}
with $V$ uniform in the unitary group of degree $N$. This composes generators from the scaling ensemble multiplicatively into a time-evolution operator, which differs from the self-similarity mapping \eqref{eq:mobius}
governing the maximally efficient scaling flow. For this process, we find again $\gamma_0=1/2$, while $\gamma_1=1-N^{-1}+O(N^{-2})$ just falls short of the chaos bound stated above at any finite value of $N$. 

\begin{figure}[ht!]
    \includegraphics[width=\columnwidth]{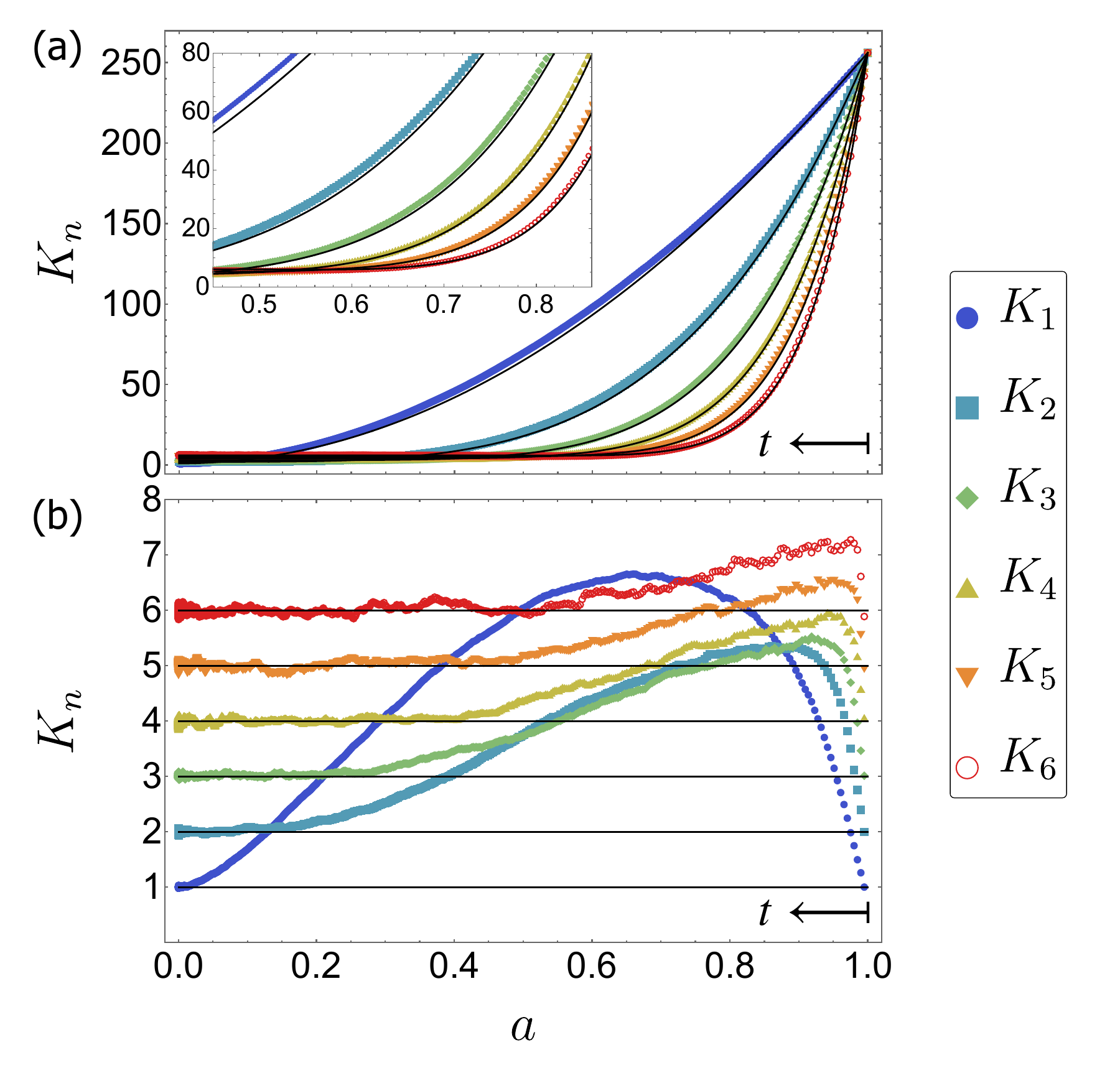}
\caption{Scaling analysis of scrambling in the Cauchy process generated by Eq.~\eqref{eq:uCauchy} ($N=16$, $dt=0.01$, $10^4$ realizations), in analogy to Fig.~\ref{fig:stochastic}.
The SFF departs from the analytical scaling predictions (black curves), which now serve as lower bounds (a), and no longer display self-similar statistics along the flow, as verified by unfolding the spectrum to uniform density (b).}
\label{fig:Cauchy}
\end{figure}

Figure~\ref{fig:Cauchy} reports the full scaling analysis of the higher-order SSFs, where we exploit the fact that the Cauchy process shares the same mean level density as the scaling theory so that no unfolding is needed (see Appendix~\ref{app:a}). 
As shown in panel (a), this now shows clear deviations from our expectations for a maximally efficient process, with the results rising significantly above the bounds \eqref{eq:StroboscopicFFResult} set by the scaling theory plotted as a function of the scaling parameter $a$. 
Panel (b) shows that these deviations persist when the spectrum is unfolded to a uniform density, which reveals that the spectral correlations are not self-similar along the scrambling dynamics. 
Therefore the framework established in this paper allows us to distinguish the effects of efficient but yet incomplete scrambling to a time $T$, captured by a finite value of $a$---already present in Figs.~\ref{fig:Stroboscopic} and \ref{fig:stochastic}---and intrinsically inefficient scrambling, captured by the departure from the scaling bounds.

\section{The out-of-time-ordered correlator} \label{sec:otoc}
As introduced in Section~\ref{sec:background}, the out-of-time-ordered correlator (OTOC) is a typical diagnostic tool that characterizes quantum-dynamical scrambling in terms of dynamical correlations. 
Specifically, the OTOC probes the spread of a time-evolved Heisenberg operator $W(t)=U^\dag(t) w U(t)$ with an operator $V(0)=v$ at an earlier time and quantifies how small perturbations spread over a system using the squared commutator 
\begin{equation} \label{eq:sqcommutator}
     \mathcal{O}(t) = \left< [W (t),V(0)]^\dag [W (t),V(0)] \right>, 
\end{equation}
where we evaluate the expectation value $\langle \cdot \rangle$ in the infinite-temperature state. 
For completeness, we consider the OTOC of a time-evolved operator with itself $v=w$, with $v$ taken to be from the GUE and scaled such that $\overline{v^2}=\openone$. Upon ensemble averaging, the OTOC takes the form {\cite{Kalsi2024HierarchicalUniversal}
\begin{align} \label{eq:otoc}
    \overline{\mathcal{O}(t)} &= 2 N \left( 1+ \frac{ {K}_{1}(t)}{N^2} - \frac{2 \mathcal{K}}{N^3}\right),
\end{align}
where we have already derived the expression for $K_n$ for all $n$ (Eq.~\eqref{eq:StroboscopicFFResult}) and must now evaluate the additional correlator 
\begin{equation} \label{eq:K112}\mathcal{K}=\overline{(\mathrm{tr}\,U(t))^2\,\mathrm{tr}\,\left[U^\dag(t)\right]^2 }
\end{equation}
in the scaling ensemble \eqref{eq:PoissonKernel}. 

\begin{widetext}
Working through this calculation in a similar fashion to the calculation of the SFF presented in Section~\ref{sec:sffderive} (see Appendix~\ref{app:d}), we arrive at the OTOC in the scaling theory,
\begin{align} \label{eq:otocresult}
    \overline{\mathcal{O}(t)}={}& 2 N \left(1+\frac{a^2 N^2-a^{2 N}+1}{N^2} 
    -\frac{2 \left(a^{2 N}-a^{2 N+4}+a^6 N\right)}{a^2 N}\right).  
\end{align}
\end{widetext}
As depicted in Figure~\ref{fig:otoc}, we find good agreement between the analytical result for the OTOC in the scaling theory \eqref{eq:otocresult} and numerical data for the Poisson kernel, valid for $a=e^{-t/2}$ .

\begin{figure}[t]
    \includegraphics[width=\columnwidth]{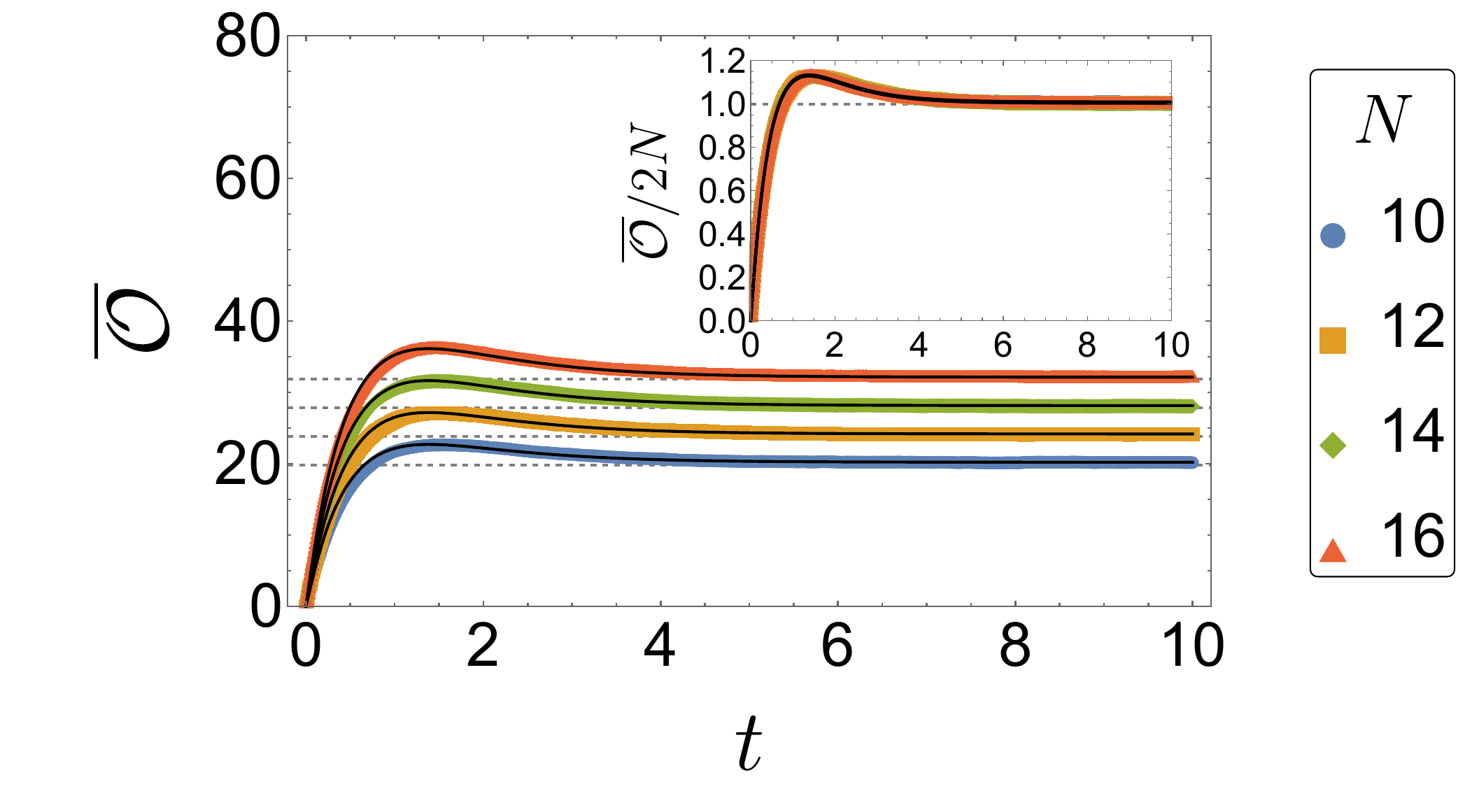}
\caption{Comparison of the analytical result for the OTOC $\mathcal{O}(t)$ in the scaling ensemble given by Eq.~\eqref{eq:otocresult} (black curves) and numerical sampling of the scaling ensemble \eqref{eq:PoissonKernel} ($dt=0.01$, $10^4$ realizations) for various system sizes $N$. The inset shows the OTOC $\mathcal{O}(t)$ under normalization by $2N$, which then tends to unity for long times.}
\label{fig:otoc}
\end{figure}

\section{Conclusions and Outlook} \label{sec:conclusions}
In summary, we developed a quantitatively predictive single-parameter scaling theory of maximally chaotic scrambling dynamics, which embodies self-similarity of the spectral correlations along the whole process. The theory is amenable to a complete analytical treatment, delivering bounds for the decay of the spectral correlations on all scales. These bounds are tightly met by Dyson's Brownian motion, which illuminates the physical content of the scaling theory and underlines the privileged nature of this paradigmatic RMT process. Signatures of inefficient or incomplete scrambling in other scenarios are captured sensitively, revealing, for instance, that the purely exponential decay of spectral correlations observed by a wide class of scrambling models is in itself not a sufficient signature of maximally chaotic scrambling. Within the scaling theory, we are further able to compute the OTOC, which can be expressed as a combination of the spectral correlation functions.

The scaling theory also concretizes deeper conceptual features of general scrambling dynamics, such as the intimate link of short-time scrambling and long-time ergodicity enforced by the unitarity of this process.
As chaotic scrambling is a fundamental tenet of complex quantum-matter phenomenology, this approach transfers to a wide range of physical domains \cite{Kalsi2024HierarchicalUniversal}. 
Interesting extensions would include the consideration of constrained dynamics, such as those obtained from symmetries placing systems into different universality classes, as well as an examination of the extremal statistics at the spectral edge.

\begin{acknowledgments}
We gratefully thank Amos Chan for his comments on the manuscript. This research was funded by EPSRC via Grant No. EP/T518037/1.
All relevant data present in this publication can be accessed at \cite{data}.
\end{acknowledgments}

\onecolumn
\appendix
\section{Mean density of states\label{app:a}}
In the main text, we give the derivation of the scaling mean density of states. In order to obtain the mean density of eigenvalues $\lambda_l\equiv\exp(i\phi_l)$ of the 
unitary time-evolution operator $U$ in the other processes, we follow a similar argument: first expressing it in terms of the moments $A_n=N^{-1}\overline{\mathrm{tr}\,U^n}$, 
and then applying  Eq.~\eqref{eq:dos}. 
Recall that in the scaling theory, $A_n = a^n$ follows directly by expanding Eq.~\eqref{eq:PoissonKernel} into a geometric series, and using the CUE average $\overline{V^m} =\delta_{0m}\openone$. 
Summation of the series \eqref{eq:dos} then delivers the scaling mean density of states \eqref{eq:scalingdos}.

We now treat the other scenarios. In the Cauchy process, we can apply the same expansion to the generator \eqref{eq:uCauchy}, so that over each time step the moments update as $A_n \rightarrow  (1-dt)^{n/2} A_n$. In the continuum limit $dt\to 0$, we then obtain $A_n = \exp(-nt/2)\equiv a^n(t)$, where the scaling parameter $a(t)=\exp(-t/2)$ now  explicitly depends on time. In terms of this parameter, the mean density of states then takes the same analytical form \eqref{eq:scalingdos} as in the scaling theory itself, while as a function of time it can be written as
\begin{equation}
\label{eq:cauchydos}
\rho(\phi)= \frac{1}{2\pi}  \frac{\sinh{t/2}}{(\cosh{t/2}-\cos{\phi})}. 
\end{equation}
This matches numerical sampling at various times in the evolution, as shown in Fig.~\ref{fig:CauchyDOS}.
\begin{figure}[t]
    \centering
    \includegraphics[width=0.5\linewidth]{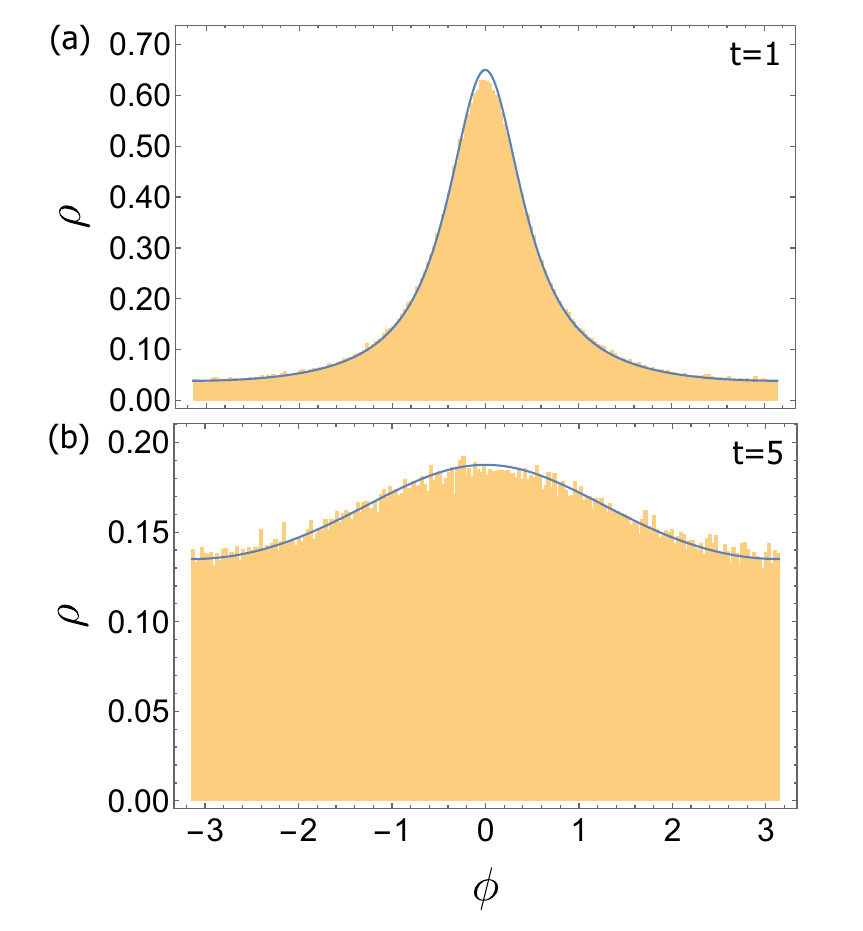}
    \caption{Comparison of the analytical density of states \eqref{eq:cauchydos} for the Cauchy process (curves) and data from numerical sampling the  process (histograms) at (a) $t=1$, and (b) $t=5$. Data generated over $10^4$ realizations for matrix dimension $N=32$.  }
    \label{fig:CauchyDOS}
\end{figure}

In the Dyson Brownian Motion (DBM) process, we perform the Gaussian averages in the generator \eqref{eq:uWiener} in the large-$N$ limit to obtain the recursive differential equations
$dA_n/dt=-(n/2)A_n-\sum_{l=1}^{n-1} (n/2)A_l A_{n-l}$, which are initialized by 
$A_1(t)\equiv a(t)=\exp(-t/2)$ and $A_n(0)=1$.
The explicit time dependence of the moments then follow as
\begin{align}
 A_n&=\exp(-nt/2)\sum_{m=0}^{n-1}\frac{(n-1)!}{m!(m+1)!(n-1-m)!}(-t n)^{m}
 \nonumber\\
 &=\exp(-nt/2) {}_1F_{1}(1-n,2,nt),
 \label{eq:anwigner}
\end{align}
 with a hypergeometric function ${}_1F_{1}$. As illustrated in  Fig.~\ref{fig:DBMDOS}, for intermediate times
the series \eqref{eq:dos} converges numerically to a shape that
resembles the Wigner semicircle, while for later times it describes the flattening out to a uniform distribution. The figure also illustrates that in all cases, the derived mean density of states conforms well to the empirical distribution obtained by random sampling of the ensembles.
\begin{figure}[t]
    \centering
    \includegraphics[width=0.5\linewidth]{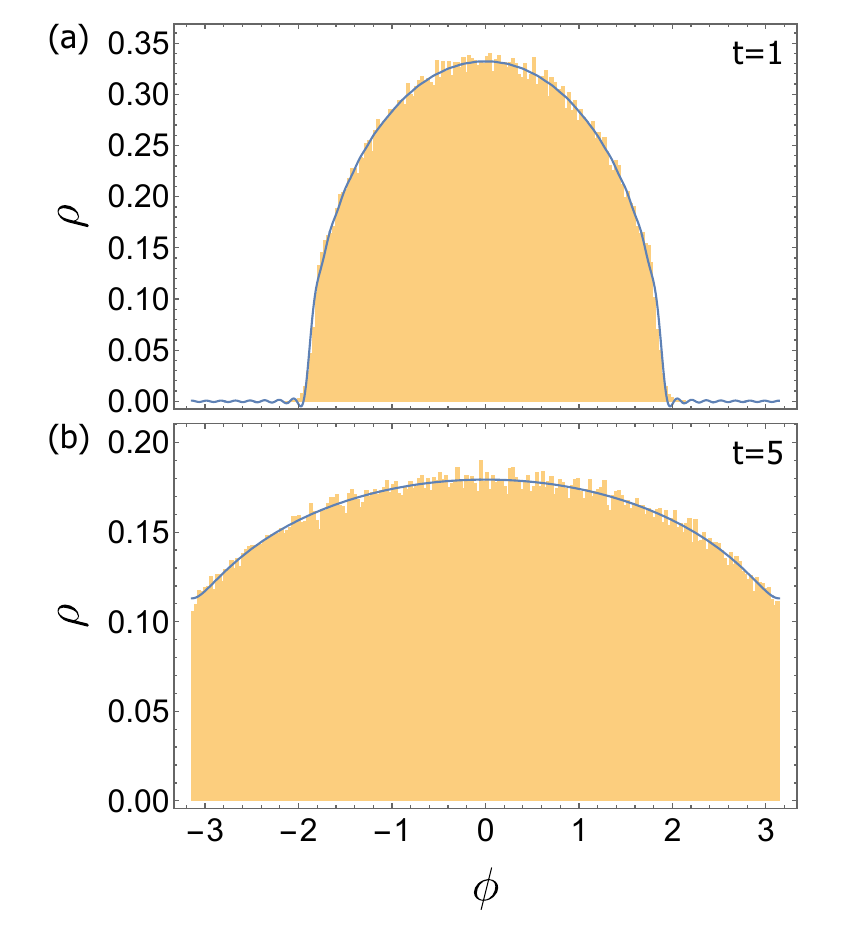}
    \caption{Analogous to Fig.~\ref{fig:CauchyDOS}, but for the Dyson Brownian motion process, where the analytical density of states is obtained by substituting the coefficients Eq.~\eqref{eq:anwigner} into Eq.~\eqref{eq:dos}.
    }
    \label{fig:DBMDOS}
\end{figure}

\section{Unfolding procedures \label{app:b}}
Our unfolding procedures are based on the direct algebraic relation of the Poisson kernel \eqref{eq:PoissonKernel}
to the CUE \cite{Savin2001}, in which the eigenvalues are distributed uniformly on the unit circle.
\begin{enumerate}
    \item For any pair of unitary matrices $U$, $V$ related by $U=(a\openone+V)/(\openone+aV)$, the 
     eigenvalues $\mu_l=\exp(i \psi_l)$ of $V$ determine the eigenvalues 
     \begin{equation}
     \label{eq:lambdamap}
     \lambda_l=(a+\mu_l)/(1+a \mu_l) = \exp(i \phi_l)
     \end{equation}
     of $U$. This is given as Eq.~\eqref{eq:lambdamapping} in the main text.
    \item If the eigenvalues $\mu_l$ are obtained from matrices $V$ of the CUE, then $\lambda_l$ are eigenvalues distributed as in the Poisson kernel, with the scaling mean density of states \eqref{eq:scalingdos}.
    \item This transformation can be inverted to translate eigenvalues $\lambda_l$ distributed with the scaling mean density of states \eqref{eq:scalingdos} into uniformly distributed eigenvalues  
    \begin{equation}
    \label{eq:mumapappendix}
  \mu_l =(a-\lambda_l)/(a \lambda_l-1),
        \end{equation}
    which is given in main text as Eq.~\eqref{eq:mumap}.     
\end{enumerate}

We can utilize these relations directly to unfold the eigenvalues in the Poisson kernel into a uniform density of states, by applying the transformation \eqref{eq:mumapappendix} individually to all eigenvalues. 
Furthermore, as the density of states \eqref{eq:cauchydos} in the Cauchy process coincides with the scaling density of states
\eqref{eq:scalingdos} of the Poisson kernel with $a(t)=\exp(-t/2)$, we can unfold the spectra from this process by the same transformation \eqref{eq:mumapappendix}. Finally, no unfolding is required to compare the original spectral statistics in these two cases for a given value of $a$.
As shown in the main text, despite this agreement of the mean density of states, the spectral fluctuations captured by the instantaneous spectral form factor (SFF) $K_n$ differ between these ensembles, both before and after unfolding.

In the DBM process, we utilize the integrated density of states
\begin{equation}
    \psi_l = \int_{min\{\phi\}}^{\phi_l} \rho(\phi)\,d\phi
\end{equation}
to unfold the eigenphases at each incremental step in the generated dynamics  to a uniform distribution. We then apply the transformation \eqref{eq:lambdamap} to unfold these eigenvalues to the scaling density of states \eqref{eq:scalingdos}.
As shown in the main text for this process, the spectral fluctuations captured by the SFF then agree with the scaling predictions \eqref{eq:StroboscopicFFResult}.

\section{Unitarily invariant processes\label{app:c}}
In this section, we derive the exponential decay laws 
\begin{align}
    a(t) &= \exp(-\gamma_0t),
    \label{eq:adecay}\\
    {K_1(t)}&=(N^2-1)\exp(-\gamma_1 t)+1,
\label{eq:kdecay}
\end{align}
of the scaling parameter and first-order SFF in unitarily invariant processes, along with their respective decay rates \eqref{eq:gamma0} and \eqref{eq:gamma1}. These expressions hold for any process in which the ensemble of stochastic generators $u(t;dt)$ are invariant under the replacement $u(t;dt)\to W^\dagger u(t;dt) W$ with arbitrary fixed matrix $W$.
The  decay laws can then be derived by performing an auxiliary average over a suitable matrix ensemble of $W$, for which we choose the CUE.

We start by considering the evolution of the scaling parameter over one incremental time step,
\begin{align}
    a(t+dt) &=N^{-1}\overline{\mathrm{tr}\,U(t+dt)}=N^{-1}\overline{\mathrm{tr}\,u(t;dt)U(t)}
    \\
    &=N^{-1}\overline{\mathrm{tr}\,W^\dagger u(t;dt)WU(t)},
\end{align}
where the last line invokes the stated invariance condition.
We next introduce the diagonalized forms $U(t)=XDX^\dagger$, $u(t;dt)=xdx^\dagger$, where $D$ and $d$ contain the eigenvalues of these matrices, while $X$ and $x$ are unitary matrices formed by the corresponding eigenvectors.
This gives 
\begin{align}
    a(t+dt) &=N^{-1}\overline{\mathrm{tr}\,w^\dagger  d wD}=N^{-1}\overline{\sum_{lm} |w_{lm}|^2  d_l D_m}
\end{align}
with  the combined unitary matrix $w= x^\dagger WX$, which inherits the  CUE distribution from $W$. 
The auxiliary average over this matrix then follows from a well-known geometric argument, which we recapitulate here for completeness.
As $w$ is a unitary matrix of dimension $N$, 
its columns (or rows) form an orthonormal
basis in $\mathbb{C}^N$,
so that by normalization and permutation symmetry of the basis indices
\begin{align} \label{eq:normalization}
    \sum_{l}|w_{lm}|^2 = 1 \rightarrow
    \overline{\sum_{l} |w_{lm}|^2} = 1 \rightarrow
    \overline{|w_{lm}|^2} = 1/N.
\end{align}
As a result,
\begin{equation}
 a(t+dt) =N^{-2}\overline{\sum_{lm} d_l D_m}=N^{-1}\overline{\mathrm{tr}\, u(t;dt)}\,a(t)
 \end{equation}
factorizes, corresponding to an
incremental change 
\begin{align}
da(t)
&=\frac{\overline{ \mathrm{tr}\, u(t;dt)}\,-N}{N} \, a(t).
\end{align}
For $dt \rightarrow 0$, this yields the universal exponential decay law
\eqref{eq:adecay},
where the decay constant 
\begin{equation}
\gamma_0=\lim_{dt\to 0}(dt\,N)^{-1}(N-\overline{\mathrm{tr}\, u(t;dt)}),
\label{eq:gamma0app}
\end{equation}
recovers Eq.~\eqref{eq:gamma0}.

Analogously, we can write the increment of the form factor as
\begin{align}
{K_1(t+dt)}=& \overline{|\,\mathrm{tr}\, u(t;dt)  U(t)|^2} =\overline{|\,\mathrm{tr}\, w^\dagger \, d \, w D|^2}
\nonumber\\=& \sum_{l,m,s,p} \overline{d_{l} D_{m} D^*_{p} d^*_{s} |w_{lm}|^2 \, |w_{sp}|^2}.
\end{align}
In the auxiliary average over $w$, we then split terms according to the index combinations
\begin{align}
  \overline{|w_{lm}|^4} &= \frac{2}{N(N+1)},\nonumber\\
  \overline{|w_{lm}|^2 |w_{lp}|^2}&=
  \overline{|w_{ml}|^2 |w_{pl}|^2} = \frac{1}{N(N+1)} \quad(m\neq p),
  \nonumber\\
         \overline{|w_{lm}|^2 |w_{sp}|^2}& = \frac{1}{N^2-1} \quad(l\neq s,m\neq p).
         \label{eq:ijkl}
\end{align}
This sums up to
\begin{align}
{K_1(t+dt)}
&=\frac{\overline{|\,\mathrm{tr}\, u|^2} \,\mathrm{tr}\, U^\dagger U + \overline{\,\mathrm{tr}\, u^\dagger u} \,|\mathrm{tr}\, U|^2}{N(N+1)} +\frac{(\overline{\,\mathrm{tr}\, u^\dagger u} -\overline{|\,\mathrm{tr}\, u|^2})(\,\mathrm{tr}\, U^\dagger U-|\,\mathrm{tr}\, U|^2)}{N^2-1}
\nonumber\\
&={K_1(t)}+\frac{N^2-\overline{|\,\mathrm{tr}\, u(t;dt)|^2}}{N^2-1}(1-K_1(t)),
\end{align}
where we momentarily suppressed the time arguments of $u(t;dt)$ and $U(t)$, and then used the unitarity of these matrices. This recovers Eq.~\eqref{eq:Invariance} in the main text, and in the continuum limit $dt\to 0$ results in the exponential decay law
\eqref{eq:kdecay},
where the constant
\begin{equation}
\gamma_1=\lim_{dt\to 0}dt^{-1}(N^2-\overline{|\,\mathrm{tr}\, u(t;dt)|^2})/(N^2-1)
\label{eq:gamma1app}
\end{equation}
is in agreement with Eq.~\eqref{eq:gamma1}.

\subsection{Application to Dyson's Brownian motion}
In the DBM process, the generators \eqref{eq:uWiener} are expressed in terms of Hamiltonians $H$ from the Gaussian unitary ensemble, satisfying
\begin{equation}
\overline{H_{lm}}=0,\quad \overline{H_{kl}H_{mn}}=N^{-1}\delta_{kn}\delta_{lm},
\end{equation}
implying $\overline{H^2}=\openone$.
For the determination of the rates $\gamma_0$ and $\gamma_1$ in the continuum limit, we can expand the generator
as 
\begin{equation}
u(t;dt)=\openone-i\sqrt{dt}H-dt H^2/2,
\end{equation}
upon which
\begin{align}
&\overline{\mathrm{tr}\,u(t;dt)}=N(1-dt/2),
\\
 &\overline{|\, \mathrm{tr}\,u(t;dt)|^2} =N^2-(N^2-1)dt.
\end{align}
Equations \eqref{eq:gamma0} and \eqref{eq:gamma1} in the main text (replicated above as Eqs.~\eqref{eq:gamma0app} and \eqref{eq:gamma1app}) then deliver the decay rates
\begin{equation}
\gamma_0=1/2,\quad\gamma_1=1.
\end{equation}

For completeness, we verify these decays by explicitly averaging in the ensemble, without using the unitary invariance. 
For the scaling parameter, we obtain
\begin{align}
a(t+dt)&=a(t)-i\sqrt{dt}N^{-1}\overline{\mathrm{tr}\,HU(t)}-\frac{dt}{2}a(t)
\nonumber\\
&=(1-dt/2)a(t),
\end{align}
such that incrementally,
\begin{equation}
    \frac{da}{dt}  = -\frac{1}{2} a(t) .
\end{equation}
From the initial condition $a(0)=1$, we then recover the exponential decay $a(t) = e^{-t/2}$ with decay constant $\gamma_0=1/2$. 

For the first-order SFF, we arrive at
\begin{align}
   {K_1(t+dt)} &= {K_1(t)} + dt \overline{\,\mathrm{tr}\,HU^\dagger(t)\,\mathrm{tr}\,HU(t)} - \frac{dt}{2} (\overline{\,\mathrm{tr}\,H^2U^\dagger(t)\, \mathrm{tr}\,U(t)} + \overline{\,\mathrm{tr}\,U^\dagger(t)\,\mathrm{tr}\,U(t)H^2}) \nonumber \\
    &= K_1(t) + N^{-1}dt\overline{\,\mathrm{tr}\,U^\dagger(t)U(t)} -dtK_1(t) \nonumber\\
    &= K_1(t) + (1-K_1(t))dt.
\end{align}
Therefore, incrementally,
\begin{equation}
    \frac{d}{dt}K_1(t)=1-K_1(t),
\end{equation}
which for the initial condition $K_1(0) = \overline{|\, \mathrm{tr}\, \openone|^2} = N^2$  is indeed solved by an exponential decay
\begin{equation}
    K_1(t) = (N^2-1)e^{-t}+1, 
\end{equation}
with decay constant $\gamma_1=1$.

\subsection{Application to the Cauchy process}
In the  Cauchy process, we can exploit the correspondence between the generators
\eqref{eq:uCauchy} and the Poisson kernel \eqref{eq:PoissonKernel} when $a=\sqrt{1-dt}$, now taken with $dt\to 0$ to obtain matrices close to the identity that then are composed multiplicatively.
Transferring Eq.~\eqref{eq:aapp} to the setting of these generators, we find 
\begin{equation}
\overline{u(t;dt)}=\sqrt{1-dt}\openone.
\end{equation} 
Equation \eqref{eq:gamma0} then determines the decay constant $\gamma_0=1/2$, which matches with the decay rate in the DBM.

The determination of the decay rate $\gamma_1$ for the first-order SFF $K_1(t)$ is considerably more involved.
We start by expanding
\begin{equation}
u(t;dt)=\frac{\beta\openone+V}{\beta V+\openone} = (\beta\openone+V)\sum_{n=0}^\infty(-\beta V)^n
\end{equation}
with $\beta = \sqrt{1-dt}$, so that
\begin{align}
    \overline{|\, \mathrm{tr}\,u(t;dt)|^2} 
    &= \underbrace{\overline{\mathrm{tr}\, \left( \beta\openone\sum_{n=0}^\infty(-\beta V)^n \right)\, \mathrm{tr}\,\left( \beta\openone\sum_{m=0}^\infty(-\beta V^\dagger)^{m} \right)}}_{A}  + \underbrace{\overline{\mathrm{tr}\, \left( \beta\openone\sum_{n=0}^\infty(-\beta V)^n \right)\, \mathrm{tr}\,\left( V^\dagger \sum_{m=0}^\infty(-\beta V^\dagger)^{m} \right)}}_{B} \nonumber\\
    &+ \underbrace{\overline{\mathrm{tr}\, \left( V \sum_{n=0}^\infty(-\beta V)^n \right)\, \mathrm{tr}\,\left( \beta\openone\sum_{m=0}^\infty(-\beta V^\dagger)^{m} \right)}}_{C}  + \underbrace{\overline{\mathrm{tr}\, \left( V\sum_{n=0}^\infty(-\beta V)^n \right)\, \mathrm{tr}\,\left( V^\dagger\sum_{m=0}^\infty(-\beta V^\dagger)^{m} \right)}}_{D}
\end{align}
breaks up into four structurally similar terms.
In each of these terms, only combinations of $V$ and $V^\dagger$ raised to the same power make finite contributions to the average, where for instance
\begin{equation}
    A = \beta^2 \, \sum_{n=0}^\infty(-\beta)^{2n} \, \overline{|\, \mathrm{tr}\,V^n|^2} \equiv \sum_{n=0}^\infty{\beta^{2n+2} \, k_n}.
\end{equation}
Here, 
\begin{equation} \label{eq:formfactor}
    k_n = \overline{|\, \mathrm{tr}\,V^n|^2} =   \begin{cases} N^2\delta_{n,0} + n & \text{for }0 \leq n \leq N\\
  N & \text{for }n > N
\end{cases}
\end{equation}
is the form factor in the CUE.
Analogously, we find
\begin{align}
    B=C&=- \sum_{n=0}^\infty{\beta^{2n+2} \, k_{n+1}}, 
    \\
     D&= \sum_{n=0}^\infty{\beta^{2n} \, k_{n+1}}.
\end{align}
Combining these results, we have
\begin{align}
    \overline{|\,\mathrm{tr}\,u(t;dt)|^2}
    &= \sum_{n=0}^\infty \beta^{2n} \left(\beta^{2} \, k_n - 2\beta^2 \, k_{n+1} + \, k_{n+1} \right).
\end{align}
With the  CUE form factors \eqref{eq:formfactor}, we then obtain
\begin{align}
    \overline{|\,\mathrm{tr}\,u(t;dt)|^2}
    &= 1 - \beta^{2 N} + N^2\beta^2 \nonumber \\
    &= 1+N^2(1- dt) -(1- dt)^N,
\end{align}
where we reinstated the definition $\beta=\sqrt{1-dt}$ in the final line. 
The decay rate $\gamma_1=N/(N+1)$ follows from Eq.~\eqref{eq:gamma1} by expanding this result for small $dt$. 

In the Cauchy process, the first-order SFF therefore decays more slowly than in  maximally efficient scrambling dynamics. In Fig.~\ref{fig:Cauchy} of the main text, we extend this analysis to the higher-order SFFs, and quantify this in terms of deviations from the scaling bounds.

\section{OTOC calculation details \label{app:d}}
The calculation details of the OTOC in Section~\ref{sec:otoc} are analogous to those of the SFF presented in Section~\ref{sec:sffderive}. 
Proceeding in a similar fashion, we use the transformation \eqref{eq:lambdamapping} to express the new correlator \eqref{eq:K112} within the scaling ensemble \eqref{eq:PoissonKernel} as
\begin{align} 
     \label{eq:K112scaling}
     & \mathcal{K}=\left( \prod_r \int_0^{2\pi} \frac{d\psi_r}{2\pi} \right) \mathrm{det}(e^{i(p-q)\psi_q}) \sum_{lmn}  \left(\frac{a+e^{i\psi_l}}{1+ae^{i\psi_l}}\right) \left(\frac{a+e^{i\psi_m}}{1+ae^{i\psi_m}}\right) \left(\frac{a+e^{-i\psi_n}}{1+ae^{-i\psi_n}}\right)^2, 
\end{align}
where the indices $q,p=1,2,...,N$ again label the rows and columns of the resulting determinant. 

Pulling the integrals over $\psi_q$ into the $q$th rows of the matrix in the determinant, we consider the sum over the indices $l,m$ and $n$ and treat the different scenarios.

Each of the diagonal cases of $l = m = n$ contributes 1 to the sum, contributing $N$ overall. 
When $l=m\neq n$, the contribution to the sum is $K_2-N$ (where we have subtracted $N$ to account for the scenario $l=m=n$). Similarly, the cases $l\neq m = n$ and $l=n\neq m$ each contribute $K_1 - N$ (again remembering to avoid double counting the case $l=m=n$). 
Finally, we must consider the case $l\neq n\neq m$, where, in analogy to the SFF calculation, we must now consider a $3\times 3$ matrix determinant, so that each term gives contributions
\begin{equation}
\mathrm{det}
\left(
\begin{array}{ccc}
 a & \left(1-a^2\right) (-a)^{-l+m-1} & \left(1-a^2\right) (-a)^{-l+n-1} \\ \\[2ex]
 \left(1-a^2\right) (-a)^{l-m-1} & a & \left(1-a^2\right) (-a)^{-m+n-1} \\ \\[2ex]
 \{\left(a^2-1\right) (-a)^{-l+n-2} & \{\left(a^2-1\right) (-a)^{-m+n-2} & a^2 \\
 \times\left[a^2 (-l+n+1) \right. &  \times\left[ a^2 (-m+n+1) \right. & {} \\
 \left.\hspace{2cm} +l-n+1 \right] \}&  \left.\hspace{2cm} +m-n+1\right]\} & {} \\
\end{array}
\right).\\ \\[2ex]
\end{equation}
}

In order to determine the specific contributions, we must treat all possible orderings of the indices $l,m$ and $n$. For $l<m<n$, this $3\times 3$ matrix determinant reduces to the form 
\begin{equation}
     \mathrm{det} \begin{pmatrix}
     a & \cdot & \cdot \\
     0 & a & \cdot \\
     \cdot & \cdot & a^2
     \end{pmatrix}, 
\end{equation}
where $\cdot$ denotes some nonzero finite element. Summing over all permitted indices, this delivers the same contribution as the case $m<l<n$, \begin{equation}
     \mathrm{det} \begin{pmatrix}
     a & 0 & \cdot \\
     \cdot & a & \cdot \\
     \cdot & \cdot & a^2
     \end{pmatrix}. 
\end{equation}
Likewise, the contributions from the cases $l<n<m$, 
\begin{equation}
     \mathrm{det} \begin{pmatrix}
     a & \cdot & \cdot \\
     0 & a & 0 \\
     \cdot & 0 & a^2
     \end{pmatrix}
\end{equation}
and $m<n<l$,
\begin{equation}
    \mathrm{det} \begin{pmatrix}
     a & 0 & 0 \\
     \cdot & a & \cdot \\
     0 & \cdot & a^2
     \end{pmatrix},
\end{equation}
respectively, are the same. 

Similarly, contributions from $n<l<m$, \begin{equation}
     \mathrm{det} \begin{pmatrix}
     a & \cdot & 0 \\
     0 & a & 0 \\
     0 & 0 & a^2
     \end{pmatrix} = a^4, 
\end{equation}
and $n<m<l$, \begin{equation}
     \mathrm{det} \begin{pmatrix}
     a & 0 & 0 \\
     \cdot & a & 0 \\
     0 & 0 & a^2
     \end{pmatrix} = a^4, 
\end{equation}
reduce to the same form.

Evaluating the index summations subject to the constraints in each case, we find that the additional correlator \eqref{eq:K112} reduces to 
\begin{equation} \label{eq:STK112}
    \mathcal{K} = N^2 a^4\left(N+a^{2N-6}(1-a^4)\right),
\end{equation}
where we have used the result \eqref{eq:StroboscopicFFResult} to evaluate
\begin{align} \label{eq:STK1K2}
    K_1 ={}& 1 + a^2 N^2 - a^{2N} \nonumber \\
    K_2 ={}& 2 + a^4 N^2 - a^{-2 + 2 N} (2 a^2 + (-1 + a^2)^2 N^2). 
\end{align}

\end{document}